\documentclass[psf,proceeding paper,accept,pdftex,oneauthor]{Definitions/mdpi} 

\firstpage{1} 
\pubvolume{8}
\issuenum{1}
\articlenumber{30}
\pubyear{2023}
\copyrightyear{2023}
\externaleditor{Academic Editor: Yue Zhao}
\datepublished{ } 
\hreflink{https://doi.org/\linebreak 10.3390/psf2023008030} % If needed use \linebreak
\doinum{psf2023008030}
\pdfoutput=1 % Uncommented for upload to arXiv.org

\usepackage[per-mode=symbol]{siunitx}
\usepackage{xspace}
\usepackage{mathrsfs}

\newcommand{\BR}{\ensuremath{\mathscr{B}}\xspace}
\newcommand{\order}{\ensuremath{\mathscr{O}}}

\newcommand{\mueeecharged}{\ensuremath{\mu^+\to e^+e^-e^+}\xspace}

\newcommand{\muegammacharged}{\ensuremath{\mu^+\to e^+\gamma}\xspace}

\newcommand{\mueeenunucharged}{\ensuremath{\mu^+\to e^+e^-e^+\overline{\nu}_\mu\nu_e}\xspace}

\newcommand{\mueXcharged}{\ensuremath{\mu^+\to e^+X}\xspace}

\newcommand{\mueacharged}{\ensuremath{\mu^+\to e^+a}\xspace}

\newcommand{\aeecharged}{\ensuremath{a\to e^+e^-}\xspace}

\newcommand{\muAenunucharged}{\ensuremath{\mu^+\to A^\prime e^+\overline{\nu}_\mu\nu_e}\xspace}

\DeclareSIUnit{\muons}{\ensuremath{\text{muons}}}
\DeclareSIUnit\permille{\text{\textperthousand}}
\DeclareSIUnit{\ifb}{\femto\barn^{-1}}
\DeclareSIUnit{\iab}{\atto\barn^{-1}}

\Title{Searching for Charged Lepton Flavour Violation with Mu3e $^{\dagger}$}

\TitleCitation{Searching for Charged Lepton Flavour Violation with Mu3e}

\Author{Ann-Kathrin Perrevoort \orcidA{} on behalf of
  the Mu3e Collaboration}

\AuthorNames{Ann-Kathrin Perrevoort}

\AuthorCitation{Perrevoort, A.-K.}
\address[1]{Karlsruhe Institute of Technology, 76131 Karlsruhe, Germany; ann-kathrin.perrevoort@kit.edu}

 \firstnote{ Presented at the 23rd International Workshop on Neutrinos from Accelerators, Salt Lake City, UT, USA, 30--31 July 2022.} 
\abstract{
The observation of lepton flavour violation (LFV) in the charged lepton sector
would be an unambiguous sign of physics beyond the Standard Model (BSM), and
thus, it is the channel of choice for many BSM searches.
LFV searches in muon decays in particular benefit from the fact that muons can
be easily produced at high rates. 
There is a global effort to search for LFV at high-intensity muon sources to
which the upcoming Mu3e experiment at the Paul Scherrer Institute (PSI) will
contribute.
The Mu3e Collaboration aims to perform a background-free search for the LFV
decay \mueeecharged with an unprecedented sensitivity in the
order of $\num{e-15}$ in the first phase of operation and $\num{e-16}$ in the
final phase---an improvement over the preceding SINDRUM experiment by four
orders of magnitude.
The high muon stopping rates and low momenta of the decay electrons make high
demands on momentum and time resolution and on the data acquisition. 
The innovative experimental concept is based on a tracking detector built from
novel ultra-thin silicon pixel sensors and scintillating fibres and tiles as
well as online event reconstruction and filtering in real time. 
}

\begin{document}

%%%%%%%%%%%%%%%%%%%%%%%%%%%%%%%%%%%%%%%%%%
\section{Introduction}

\begin{figure}[H] %% [H]
\hspace{0.1cm}\includegraphics[width=4.3 cm, clip, trim= 0.7cm 0.3cm  0.7cm 0.25cm]{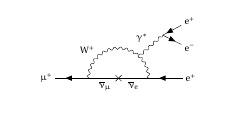}\hspace{0.2cm}
\includegraphics[width=4.3 cm, clip, trim= 0.7cm 0.35cm 0.7cm 0.25cm]{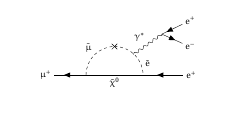}\hspace{0.2cm}
\includegraphics[width=4.3 cm, clip, trim= 0.4cm 1cm    0.4cm 0.25cm]{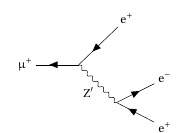}
\caption{Feynman diagrams of the \mueeecharged decay mediated via
  (\textbf{left})      neutrino mixing,
  (\textbf{centre})~supersymmetric particles,
  and (\textbf{right}) a $Z^\prime$ in models with an extended electroweak sector.
  \label{figMueee}}
\end{figure}

In the original formulation of the Standard Model (SM) of particle physics, lepton
flavour is a conserved quantity, although, this is only due to an accidental symmetry.~With the observation of neutrino mixing, it became evident that lepton flavour
is indeed not conserved in nature---at least in the neutrino sector---but so
far, lepton flavour violation in the charged lepton sector (cLFV) has eluded
observation.

cLFV processes like \mueeecharged could be mediated via neutrino mixing (see
Figure~\ref{figMueee} on the left), but they would be suppressed to branching
ratios below $\num{e-50}$ and thus far below the reach of current or upcoming
cLFV searches.
In BSM models which address for example the generation of neutrino masses or
the origin of the flavour structure, however, cLFV often occurs at observable
levels.
Any observation of cLFV would, thus, be an unambiguous sign of BSM
physics.
Examples for the \mueeecharged process are shown in Figure~\ref{figMueee} in
the centre and on the right. 

Searches for cLFV in muon decays are particularly sensitive probes of BSM,
since muons can be produced at very high intensities allowing to test also
very rare processes.
For example, PSI operates regular muon beam lines with rates of
\SI{e8}{\muons\per\second}. 
There is a ongoing global effort to search for cLFV with muons in various
channels. 
The Mu3e experiment at PSI is the only experiment planned at the moment which
is going to search for \mueeecharged. 

\vspace{-8pt}   
\unskip

%%%%%%%%%%%%%%%%%%%%%%%%%%%%%%%%%%%%%%%%%%
\section{The Mu3e Experiment}

The Mu3e Collaboration aims to find or exclude the cLFV decay \mueeecharged
with a sensitivity to branching ratios in the order of \num{e-15} in phase~I
and \num{e-16} in phase~II of the experiment~\cite{Mu3eTDR}---surpassing the
current strongest limit of \mbox{$\BR(\mueeecharged)<\num{1.0e-12}$} at a 
\SI{90}{\percent} confidence level (CL) set by the SINDRUM
experiment~\cite{SINDRUM:1987nra} by four orders \mbox{of magnitude}. 

The search will be conducted free of background.
The SM background process \mbox{\mueeenunucharged} is distinguished from signal
decays solely by the momentum carried away by the undetectable neutrinos.
A further source of background stems from accidental combinations of $e^+$ and
$e^-$ from SM muon decays, photon conversion, Bhabha scattering and
misreconstructed tracks. 
This type of background can be suppressed by kinematic selections as well as
selections on the reconstructed vertex and the coincidence of the decay~products.

In addition to excellent momentum and high vertex and time resolution,
operating at high muon decay rates of \SI{1e8}{\muons\per\second} (phase~I) up
to \SI{2e9}{\muons\per\second} (phase~II) puts further demands on the detector
and data acquisition (DAQ). 

The phase~I detector is under construction and will be operated at
the Compact Muon Beamline at PSI. 
For phase~II, the detector will be upgraded and operated at the new High-Intensity Muon Beamline, which is currently being planned at PSI~\cite{HIMB}.
In the following, the phase~I experiment is discussed in further detail.

%%%%%%%%%%%%%%%%%%%%%%%%%%%%%%%%%%%%%%%%%%
\subsection{Detector Concept}

\begin{figure}[H] %% [H]
\hspace{-0.1cm}\includegraphics[width=10.35 cm]{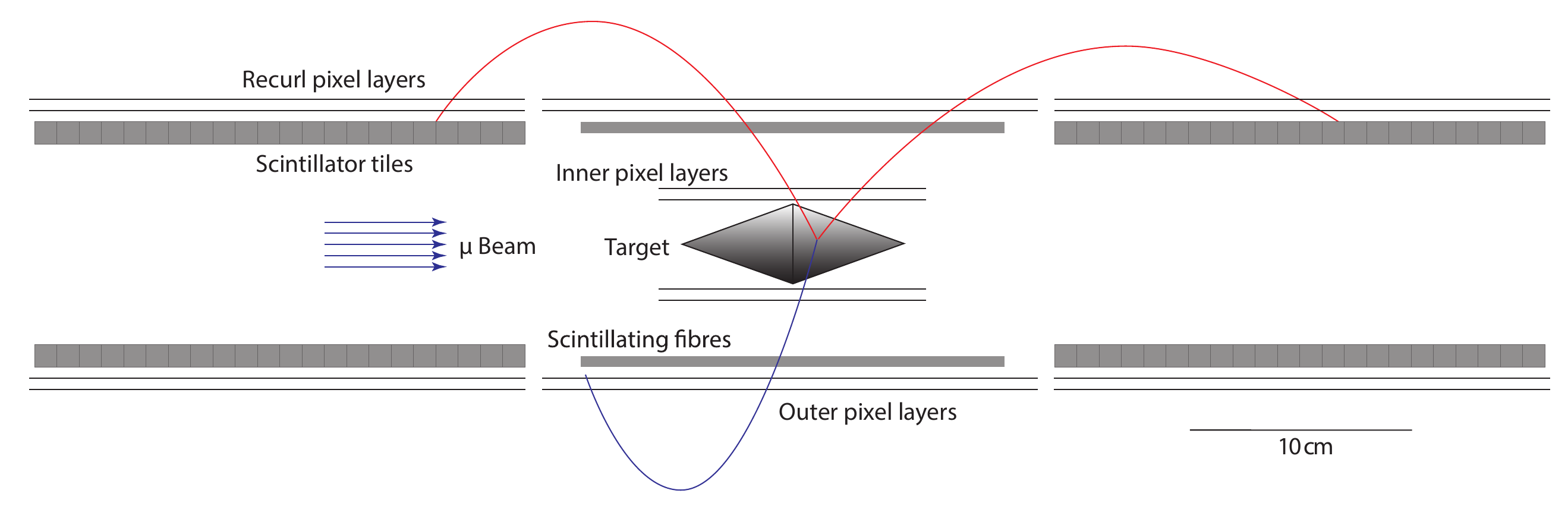}
\includegraphics[width= 3.35 cm]{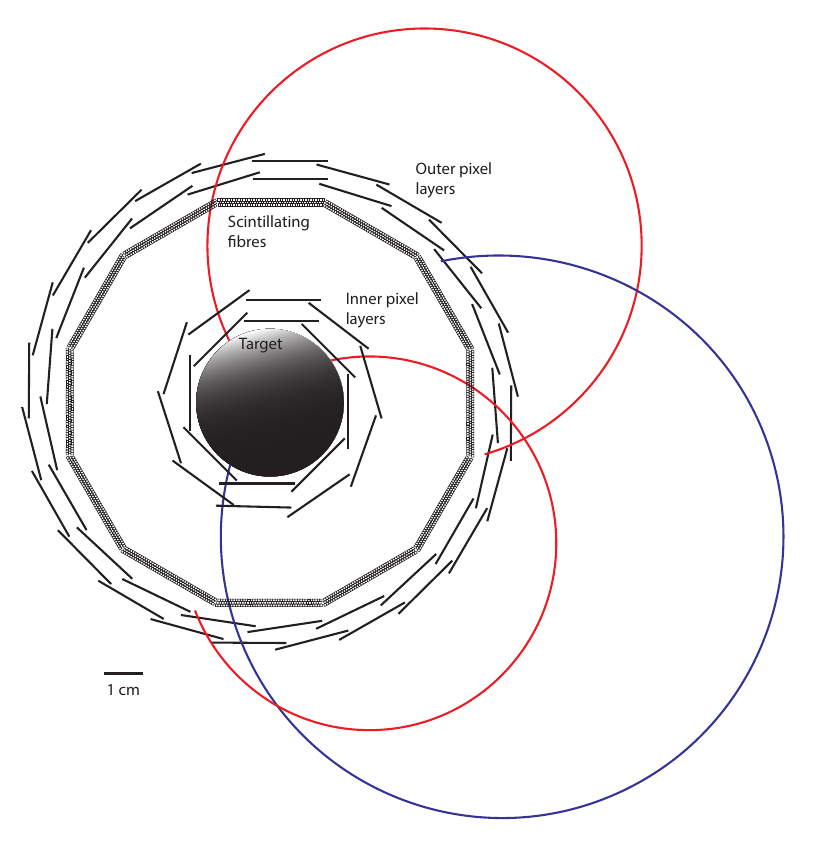}
\caption{Schematic of the Mu3e phase~I detector shown (\textbf{left}) along
  the beam axis and (\textbf{right})~transverse to the beam axis.
  A potential \mueeecharged signal decay is shown with $e^+$ trajectories
  in red and the $e^-$ trajectory in blue. \label{figDet}}
\end{figure}

The Mu3e experiment is a spectrometer placed in a \SI{1}{\tesla} solenoidal
magnetic field.
Multiple Coulomb scattering dominates the momentum resolution of the
experiment as the decay particles have momenta of only a few
\SI{10}{\mega\eV}.
For this reason, the material in the active detector volume is kept to a
minimum. 
A schematic of the Mu3e experiment is shown in Figure~\ref{figDet}. 

The $\mu^+$ beam is stopped in a thin, hollow, double-cone target built from
Mylar in the centre of the detector.
The trajectories of the decay $e^+$ and $e^-$ are measured with a
barrel-shaped, silicon pixel tracker.
Mu3e utilises \SI{50}{\micro\metre} thin pixel sensors built in the High-Voltage Monolithic Active Pixel Sensor technology~\cite{Peric:2007zz} leading
to a material amount of only \SI{0.1}{\percent} of a radiation length per
tracking layer including 
the flex-print for readout and powering and the mechanical support structures.
There are four tracking layers in the central detector part, two of which are
located close to the target.
The final prototype of pixel sensors for Mu3e---the MuPix11---has been
produced in 2022 and is currently passing the last steps of characterisation. 

In addition to the pixel tracker, a scintillating fibre detector provides a
precise timing measurement.
The fibre detector consists of three layers of \SI{250}{\micro\metre} diameter
fibres which are connected to a silicon photomultiplier column array read out
by a custom-made ASIC: the MuTRiG chip~\cite{Chen:2017qor}.
The final version of the MuTRiG has been produced in 2022 and is currently
being tested. 

The momentum resolution of the experiment is significantly improved by the
installation of so-called recurl stations upstream and downstream of the
central detector station.
Due to the bending in the magnetic field, the $e^+$ and $e^-$ produced on the
target are forced to return---\emph{recurl}---to the detector and are
measured for a second time either in the central or in the recurl stations.
Because of the large lever arm between the measurements of the outgoing and
recurling particle, scattering-induced uncertainties cancel to first order. 
The recurl stations consist of two tracking layers and scintillating tiles
for improved timing.
The scintillating tiles are read out via silicon photomultipliers and the
MuTRiG chip, which is used for the scintillating fibre detectors as well.

Components in the active detector volume are cooled with gaseous helium.
Outside the active detector volume, an additional water-cooling system is
installed in the support~structures. 

The streaming \textls[-15]{data acquisition system of Mu3e continuously proce}sses 
zero-suppressed data from all detector systems without a hardware trigger. 
Events of interest, i.e., events containing at least two $e^+$ and one $e^-$
trajectory compatible with a common vertex, are selected on the event filter
farm.
The filter farm performs fast, simplified track fits and vertex finding in
real time on Graphics Processing Units.
Raw data of events with \mueeecharged signal candidates are stored on disk for
offline analysis.
In this way, the output data rate is reduced by around a factor of 100
compared to the incoming data rate at the filter~farm.

\vspace{-3pt}   
\unskip

%%%%%%%%%%%%%%%%%%%%%%%%%%%%%%%%%%%%%%%%%%
\subsection{Sensitivity Studies}

\begin{figure}[H] %% [H]
\includegraphics[width=7.35 cm]{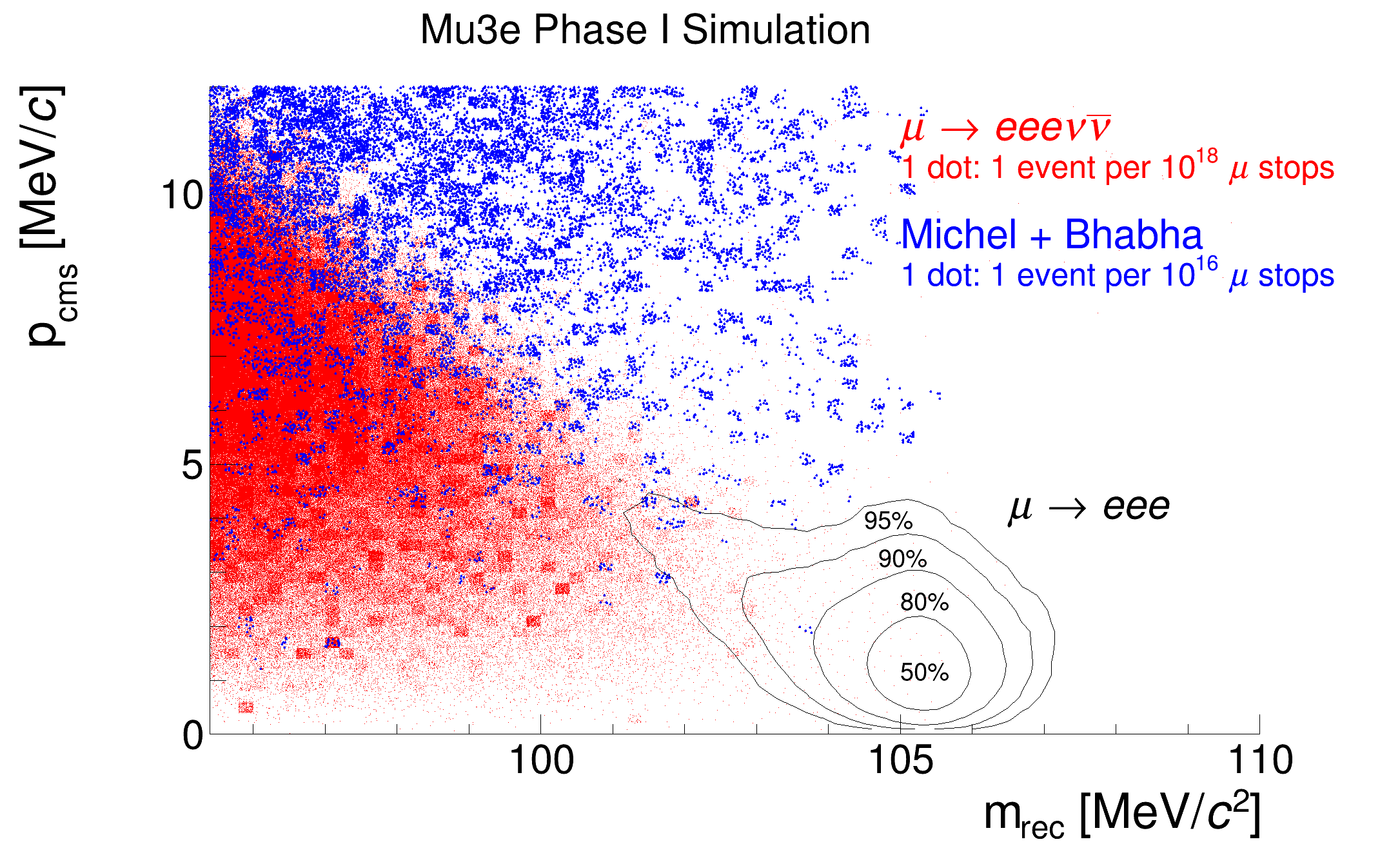}
\includegraphics[width=6.35 cm]{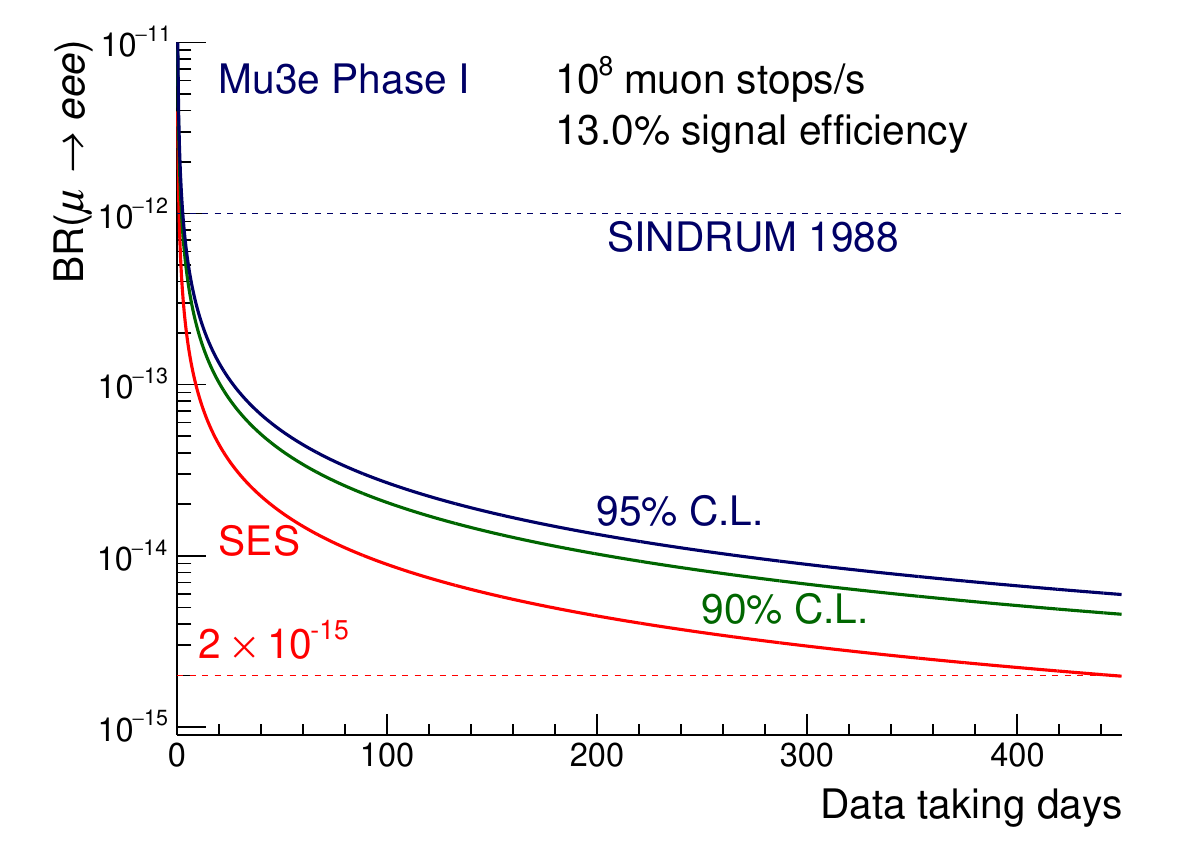}
\caption{Simulation studies of the phase~I Mu3e experiment.
  (\textbf{Left}) Distribution of simulated signal and background events.
  (\textbf{Right}) Expected sensitivity of the \mueeecharged search in phase~I.
  \label{figSensitivity}}
\end{figure}

The feasibility of a background-free search for \mueeecharged with the phase~I
Mu3e experiment up to the envisaged sensitivity has been
demonstrated with a detailed Geant4-based detector simulation.
The distribution of signal and background events in the centre-of-mass
momentum ($p_\text{cms}$) vs.\,invariant mass ($m_{eee}$) plane of the
$e^+e^-e^+$ system after kinematic, vertex and coincidence selections is shown
in Figure~\ref{figSensitivity} alongside the expected reach in
$\BR(\mueeecharged)$ in dependence of the runtime.
Branching ratios of \num{e-14} to a few \num{e-15} can be reached with 200 to
300 days of data taking. 

\begin{figure}[H] %% [H]
\includegraphics[width=4.5 cm]{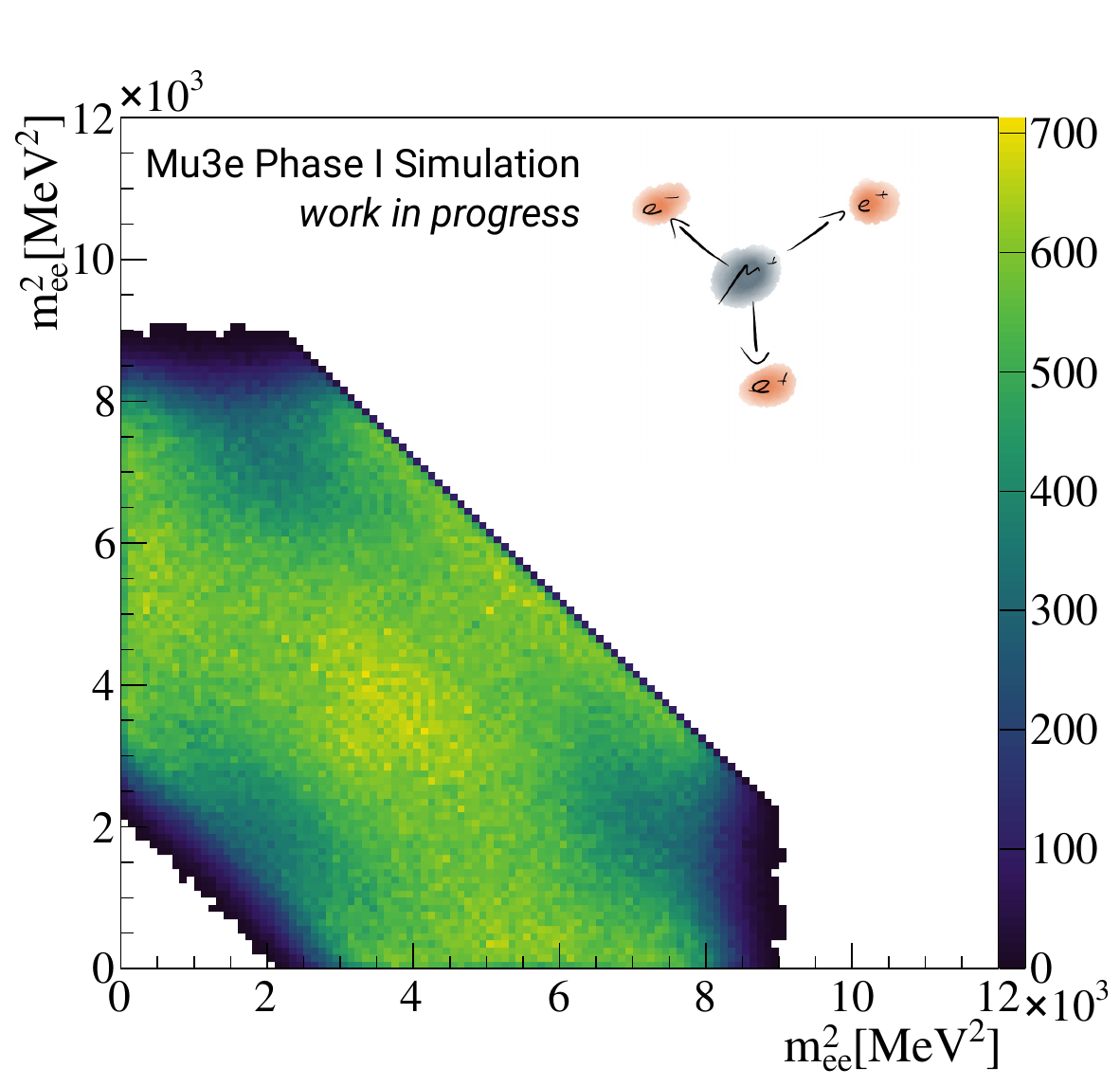}
\includegraphics[width=4.5 cm]{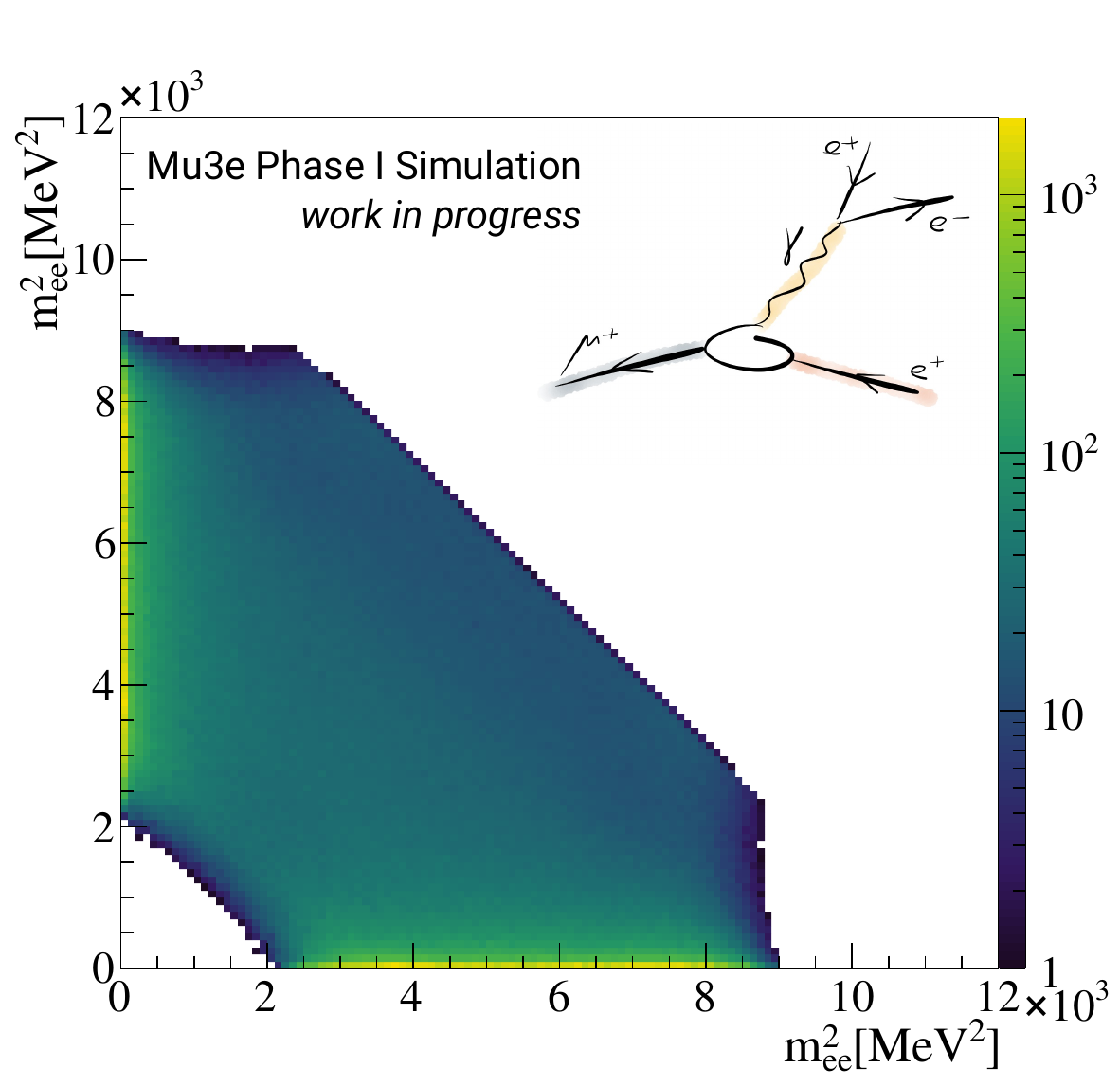}
\includegraphics[width=4.5 cm]{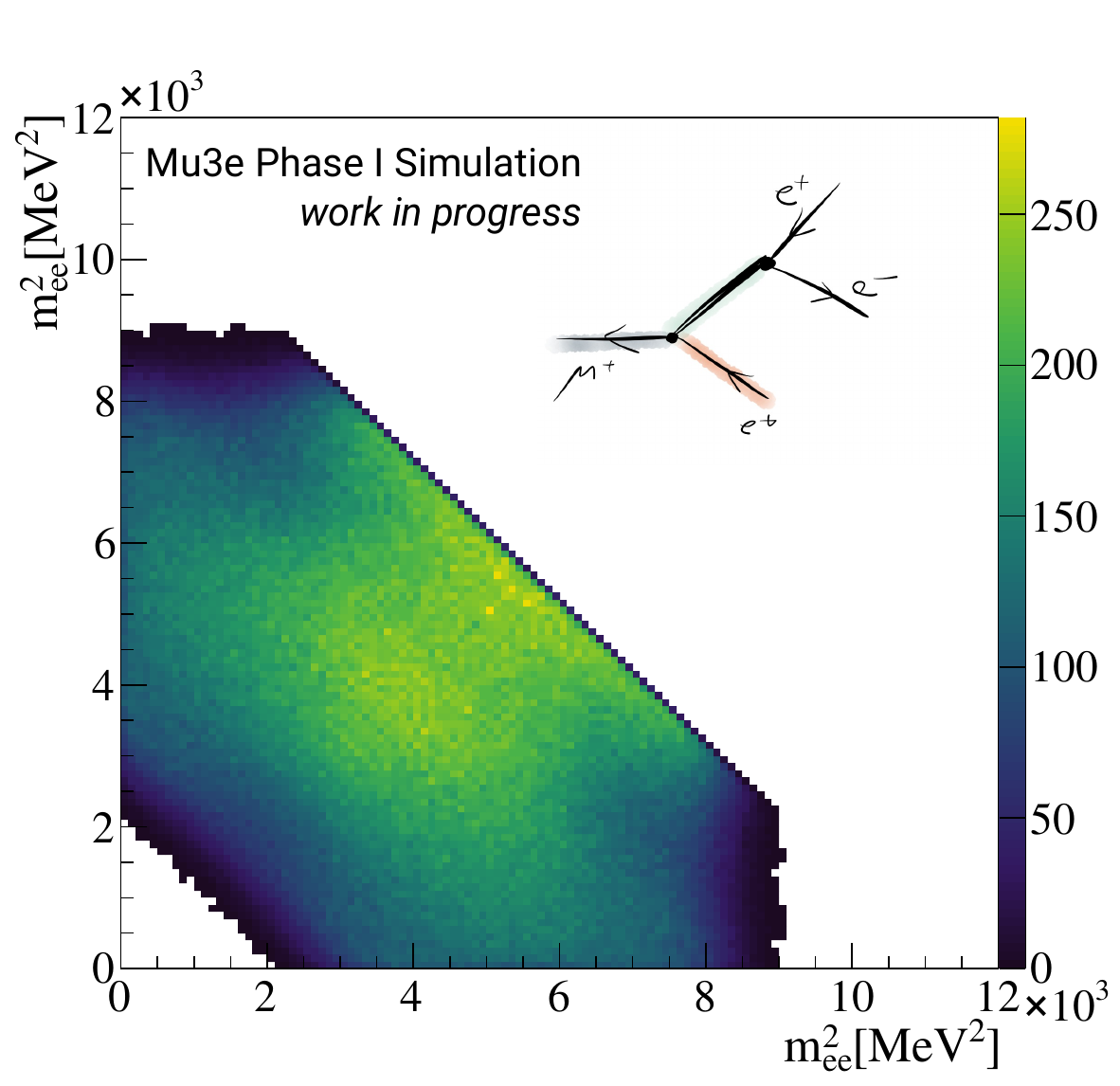}
\caption{Dalitz plots of the invariant mass of $e^+e^-$ pairs in simulated and
  reconstructed \mbox{\mueeecharged} signal decays in the phase~I Mu3e experiment
  assuming 
  (\textbf{left})   phase-space distributed decays, 
  (\textbf{centre}) an effective dipole interaction, and
  (\textbf{right}) an effective four-fermion interaction. 
  The effective Lagrangian from~\cite{Kuno:1999jp} has been deployed in this
  study. 
  \label{figEFT}}
\end{figure}

In the case of discovery and given that a sufficient number of \mueeecharged
events is observed, conclusions on the type of BSM interaction can be drawn
from the kinematics of the $e^+e^-e^+$ final state---in addition to the
interplay with observation and non-observation in searches for 
\muegammacharged and muon-to-electron conversion on nuclei.
In Figure~\ref{figEFT}, Dalitz plots of the invariant mass of the two possible
$e^+e^-$ combinations in \mueeecharged are shown assuming selected effective
operators. 

\begin{figure}[H] %% [H]
\includegraphics[width=4.4 cm]{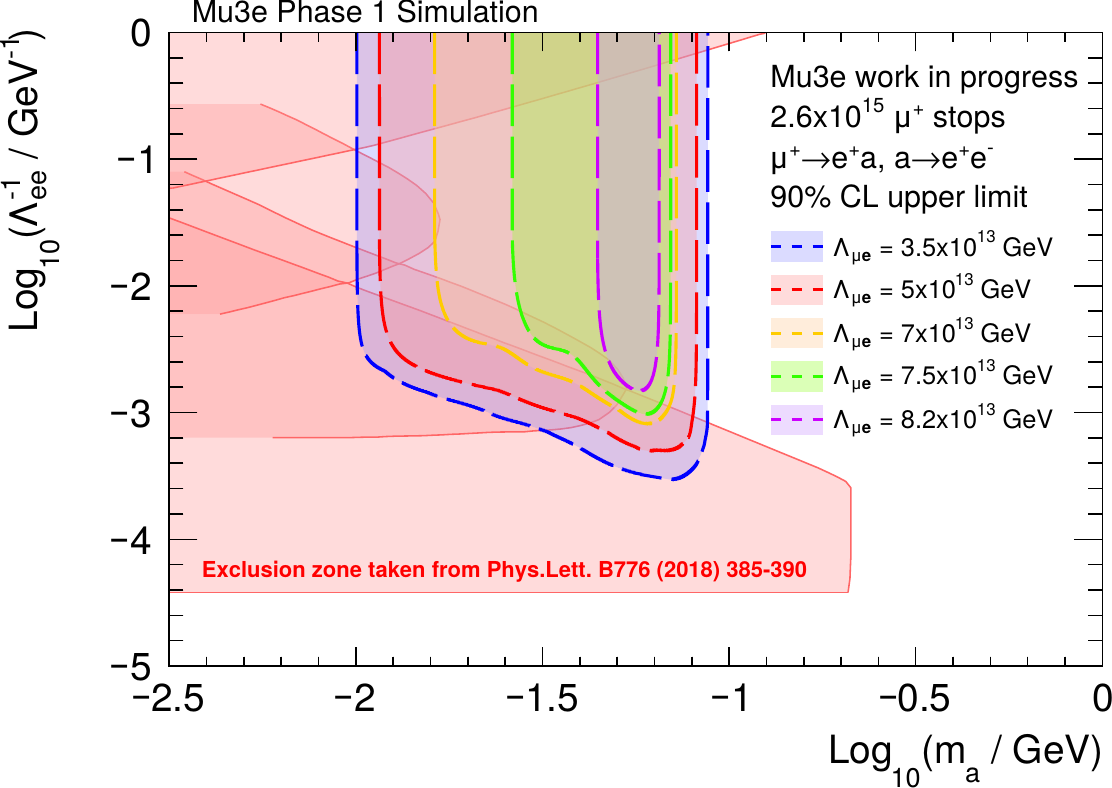}
\includegraphics[width=4.55 cm]{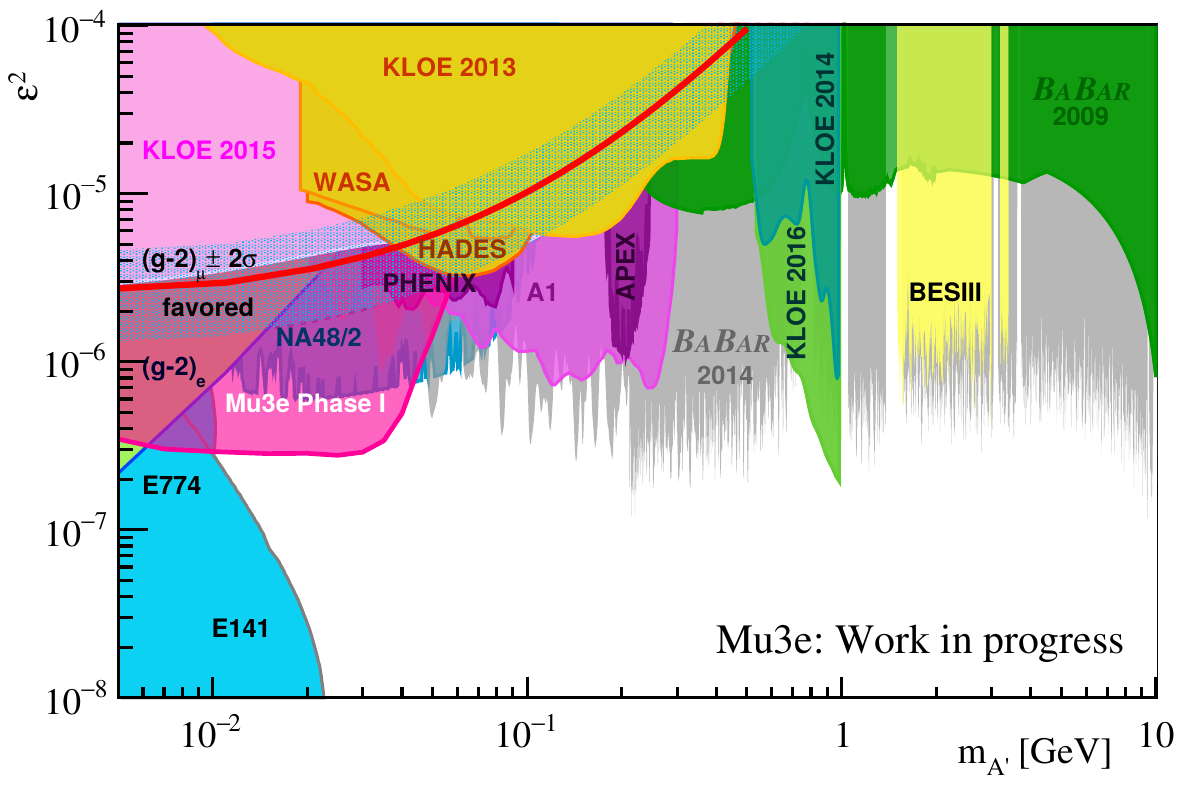}
\includegraphics[width=4.4 cm]{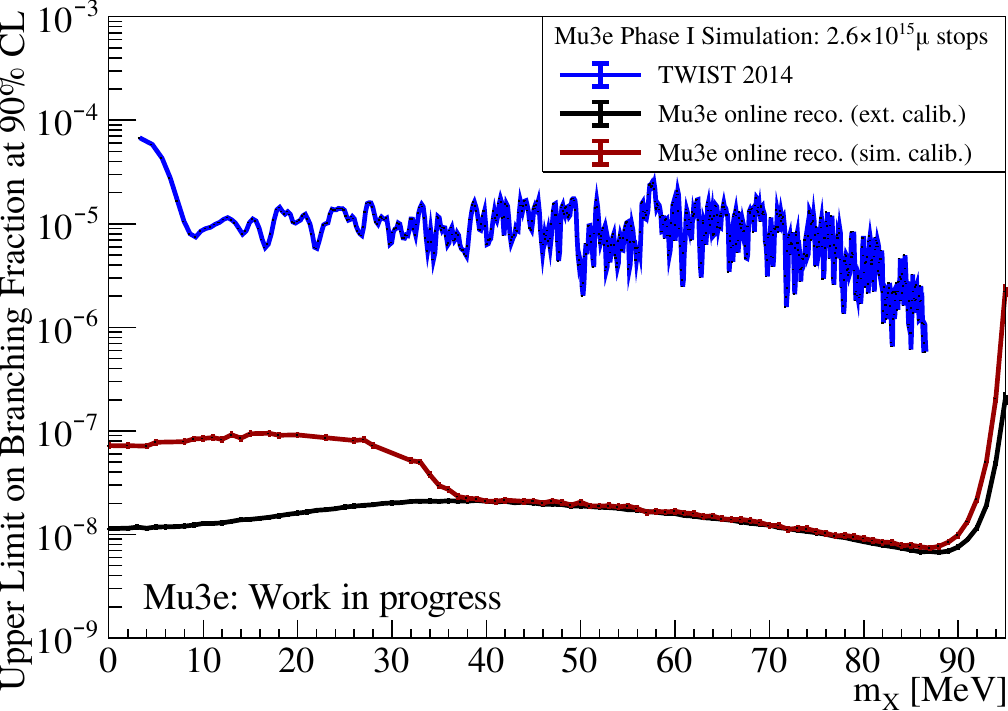}
\caption{
  Sensitivity 
  of the Mu3e phase~I experiment to certain BSM models. 
  (\textbf{Left}) Reach of a search for \mueacharged with subsequent
  \aeecharged decay in the parameter space of an axion-like $a$ as presented
  in~\cite{Heeck:2017xmg}. 
  (\textbf{Centre}) Reach of a search for promptly decaying dark photons emitted
  in muon decays.
  Lagrangian taken from~\cite{Echenard:2014lma}.
  Plot adapted from~\cite{BESIII:2017fwv}. 
  (\textbf{Left}) Sensitivity of Mu3e phase~I to the branching ratio of
  \mueXcharged compared to the current strongest limits set by
  TWIST~\cite{Bayes:2014lxz}.
  If the calibration of the total momentum scale is performed with the Michel
  spectrum, the sensitivity deteriorates at low $m_X$ (sim.\,calib.).
  Alternative calibrations are currently under investigation (ext.\,calib.).
  \label{figExotics}}
\end{figure}

As shown in Figure~\ref{figExotics} on the left, the \mueeecharged search is
also sensitive to decays of the type \mueacharged in which the particle $a$
decays within $\order(\si{\nano\second})$ to an $e^+e^-$ pair.
An example are axion-like particles as discussed in~\cite{Heeck:2017xmg}. 

\vspace{-3pt}   
\unskip

\vspace{4pt}   
%\unskip

The unprecedented data set of muon decays expected to be recorded with Mu3e can
also be exploited for other BSM searches.
Dark photons $A^\prime$ emitted in muon decays, \muAenunucharged, and promptly
decaying to an $e^+e^-$ pair can be identified in a search for a resonance in
the invariant $m_{e^+e^-}$ spectrum. 
The sensitivity of the phase~I Mu3e experiment to promptly decaying dark
photons is shown in Figure~\ref{figExotics} in the centre. 
 
Decays of the type \mueXcharged are motivated for example by familon models
which try to explain the flavour structure of the SM~\cite{Wilczek:1982rv}.
If the axion-like particle $X$ exits the detector unseen, the characteristic
signature of this decay becomes a mono-energetic $e^+$ in dependence of the
mass $m_X$ of $X$.
The Mu3e experiment is planning to implement online histograms filled with
results from track fits performed on the filter farm in which \mueXcharged
decays would show up as an excess on the smooth momentum spectrum from SM muon
decays.
The sensitivity to \mueXcharged decays of the phase~I Mu3e experiment is
expected to surpass current limits set by the TWIST experiment by two orders
of magnitude~\cite{Bayes:2014lxz} (see Figure~\ref{figExotics} on the right). 

\vspace{-3pt}   
\unskip

%%%%%%%%%%%%%%%%%%%%%%%%%%%%%%%%%%%%%%%%%%
\section{Summary and Status}

The upcoming Mu3e experiment at PSI aims to find or exclude the cLFV decay
\mbox{\mueeecharged} with an unprecedented sensitivity to branching ratios as low as
\num{e-16}. 
In a first phase, branching ratios of \num{e-14} to a few \num{e-15} can be
studied.
In addition, dark photons emitted in muon decays can be investigated with
competitive sensitivity, and current limits on \mueXcharged decays can be
surpassed by two orders of magnitude already with the phase~I Mu3e
experiment. 

Currently, the design and prototyping phase of the phase~I Mu3e experiment is
finishing and the experiment is transitioning to the production and
construction phase.
Demonstrator detector modules have been successfully operated in an
integration run in 2021 and a cosmics run in 2022. 
Commissioning and first physics data taking are expected for 2024.

%%%%%%%%%%%%%%%%%%%%%%%%%%%%%%%%%%%%%%%%%%
\vspace{6pt}

%%%%%%%%%%%%%%%%%%%%%%%%%%%%%%%%%%%%%%%%%%
\funding{The author's work is funded by the Federal Ministry of Education
and Research (BMBF) and the Baden-W\"urttemberg Ministry of Science as
part of the Excellence Strategy of the German Federal and State
Governments.  The author further acknowledges the support by the
German Research Foundation (DFG) funded Research Training Group
“Particle Physics beyond the Standard Model” (GK 1994) on previous
works on this subject. 
}

\institutionalreview{Not applicable}

\informedconsent{Not applicable}

\dataavailability{Data sharing not applicable}

\conflictsofinterest{The author declares no conflict of interest.} 

%%%%%%%%%%%%%%%%%%%%%%%%%%%%%%%%%%%%%%%%%%

\abbreviations{Abbreviations}{
The following abbreviations are used in this manuscript:\\

\noindent 
\begin{tabular}{@{}ll}
SM      & Standard Model \\
BSM     & Beyond the Standard Model\\
(c)LFV  & (charged) Lepton Flavour Violation\\
PSI     & Paul Scherrer Institute\\
CL      & Confidence Level\\
DAQ     & Data Acquisition\\
ASIC    & Application-Specific Integrated Circuit\\
MuTRiG  & Muon Timing Resolver including Gigabit-Link\\
HV-MAPS & High-Voltage Monolithic Active Pixel Sensors
\end{tabular}
}

%%%%%%%%%%%%%%%%%%%%%%%%%%%%%%%%%%%%%%%%%%
\begin{adjustwidth}{-\extralength}{0cm}

\reftitle{References}


%%%%%%%%%%%%%%%%%%%%%%%%%%%%%%%%%%%%%%%%%%
%% for journal Sci
%\reviewreports{\\
%Reviewer 1 comments and authors’ response\\
%Reviewer 2 comments and authors’ response\\
%Reviewer 3 comments and authors’ response
%}
%%%%%%%%%%%%%%%%%%%%%%%%%%%%%%%%%%%%%%%%%%
\PublishersNote{}
\end{adjustwidth}
\end{document}